\documentstyle[pre,aps]{revtex}
\tighten
\begin{document}
\draft
\twocolumn

\title{ Early stages of ramified growth in quasi-two-dimensional
electrochemical deposition}

\author{John R. de Bruyn}
\address{Department of Physics\\
Memorial University of Newfoundland\\
St. John's, Newfoundland, Canada A1B 3X7}
\date{\today}
\maketitle

\begin{abstract}

I have measured the early stages of the growth of branched metal
aggregates formed by electrochemical deposition in very thin layers.
The growth rate of spatial Fourier modes is described qualitatively by
the results of a linear stability analysis [D.P. Barkey, R.H. Muller,
and C.W. Tobias, J. Electrochem. Soc. {\bf 136}, 2207 (1989)]. The
maximum growth rate is proportional to $(I/c)^\delta$ where $I$ is the
current through the electrochemical cell, $c$ the electrolyte
concentration, and $\delta = 1.37 \pm 0.08$. Differences between my
results and the theoretical predictions suggest that electroconvection
in the electrolyte has a large influence on the instability leading to
ramified growth.

\end{abstract}


Rough surfaces develop in many systems in which the surface grows
under far-from-equilibrium conditions \cite{bg90,v92}, due to the fact
that the smooth surface is unstable to the growth of perturbations
over a range of wave numbers. This morphological instability was first
treated by Mullins and Sekerka (MS) in the context of solidifying
alloys \cite{ms}. Metal aggregates grown by electrochemical deposition
(ECD) in quasi-two-dimensional geometries display a variety of
qualitatively different branched growth morphologies
\cite{sdg86,gbcs86}, and there has been considerable study of the
basic instability of the straight electrode which leads to the initial
development of branched growth
\cite{bb62,akkf80,am84b,halsey,kkl88,BMT89,c90,pf92,kzfw92}.

To understand this instability qualitatively, assume that the cations
are transported by migration in an applied electric field to the
cathode, where they are deposited. The electric field will be uniform
along the length of a straight cathode, but will be enhanced near the
tip of a small bump.  This will lead to an increased current of
cations, and so a higher deposition rate, at the tip of the bump.
Thus the bump will grow exponentially. A similar argument applies if
the transport is diffusive. At very high spatial frequencies, however,
the straight interface is stabilized by surface tension. This picture
leads to an instability of the straight electrode analogous to the
Mullins-Sekerka instability --- perturbations within a band of wave
numbers $k$ from $k=0$ up to a cutoff wave number $k_c$ will grow,
while perturbations with $k>k_c$ will be damped out. The exponential
growth rate $\beta$ of spatial modes with wave number $k$ has the form
\cite{ms}
\begin {equation}
\beta(k) = qk(1-rk^2),
\label{MS}
\end{equation}
 with $k_c = r^{-1/2}$.

Experimentally, Kahanda et al. \cite{kzfw92} studied growth by
electrochemical deposition in the limit of very low currents. In this
case \cite{kzfw92}, the deposition of metal onto the cathode is
limited by the activation of cations in the double layer. Kahanda et
al. used a Fourier analysis technique to study the growing aggregate
front. They determined the growth rate of the Fourier modes as a
function of wave number and found semi-quantitative agreement with the
predictions of MS theory.

At higher currents, however, the deposition rate is governed by the
transport of ions in the electrolyte. A number of electrochemical
processes contribute to the ion transport and to the current
distribution near the cathode, resulting in modifications to Eq.
\ref{MS}
\cite{akkf80,am84b,halsey,BMT89,pf92}.
Barkey, Muller, and Tobias (BMT) \cite{BMT89} have performed a linear
stability analysis of a planar electrode in three-dimensional ECD,
and, at least conceptually, the results of their work should carry
over to the two-dimensional case studied here. They assume mass
transport to be due solely to diffusion. Surface tension manifests
itself as a reduction in the binding energy of deposited atoms on a
curved surface relative to a flat surface (referred to in BMT as the
capillary potential shift). The effects of electrode reactions
(kinetic overpotential) and of charge separation in the diffusion
layer close to the cathode (concentration overpotential) both act to
smooth out variations in the current density, and thus reduce the
growth rate of perturbations.  BMT found (see Eq. (41) of Ref.
\cite{BMT89}) the growth rate $\beta(k)$ to have the form
\begin{equation}
\beta = {qk (1-rk^2)\over (1+sk)}.
\label{BMT}
\end{equation}
The coefficients $q$, $r$, and $s$ depend on the properties of the
electrolyte and the deposited metal, the ion concentration, and the
current \cite{parameters}. Roughly speaking, $q$ incorporates the
destabilizing effects of mass transport, $qr$ involves the stabilizing
effects of surface tension, and $s$ is a combination of terms
involving the kinetic overpotential and the concentration
overpotential. As above, there is a range of unstable wave numbers,
$0<k<r^{-1/2}$; the MS result is recovered in the limit that $s
\rightarrow 0$.

In this Communication, I present results from experiments on the very
early stages of the ECD of aggregates of metallic copper from
solutions of CuSO$_4$, in the regime where the growth is limited by
mass transport in the electrolyte.  Using analysis techniques similar
to those of Kahanda et al. \cite{kzfw92}, I investigate the growth
rate $\beta(k)$ and find it to be qualitatively described by the
dispersion relation of BMT (Eq.
\ref{BMT}) and not by Eq. \ref{MS}.

In my experiments \cite{ld95}, two copper foil electrodes 5.1 cm long
by 0.025 cm thick, separated by roughly 23 mm, were sandwiched between
two 5.1 cm square by 0.6 cm thick glass plates and clamped together.
The space between the electrodes was filled with aqueous solutions of
CuSO$_4$ with concentrations $c$ in the range 0.02 M $<c<$ 0.5 M. A
constant current $I$ of from 0.06 mA to 10 mA was passed through the
cell, and a branched copper aggregate formed at the cathode. The cell
was illuminated from below by a diffuse white light source, and imaged
from above with a ccd video microscope. Individual images of the
aggregate were captured and digitized by a frame grabber in a personal
computer, and the growth was also recorded on video tape. The spatial
resolution of the digitized images was typically 23 $\mu$m/pixel.

My data analysis was similar to that of Kahanda et al. \cite{kzfw92}
The edge of the aggregate (the interface) was located by thresholding
the digitized image. Pixels with an intensity lower than a chosen
value were deemed to be on the aggregate. Since the aggregate is
rough, the interface determined in this way, $p(x)$, will in general
not be a single-valued function of the coordinate $x$ along the length
of the cathode. For my analysis, I formed a single-valued
approximation to the interface, $p_1(x)$, by taking the point on the
interface furthest from the original cathode position, i.e., the
highest point, at each $x$ value.  For a very heavily branched
aggregate with many ``overhangs,'' $p_1(x)$ is not a good
representation of the true interface. However, for the early times of
interest here, the aggregate, although rough, has few overhangs, and
$p_1(x)$ is very similar to $p(x)$.  The single-valued interface
function $p_1(x)$ is then Fourier transformed to give a spatial
Fourier power spectrum of the interface \cite{nr}.

Fig. \ref{interfs} shows the interface functions determined at two
different times for a particular run. Curve (a) of Fig. \ref{interfs}
shows $p(x)$ determined from a digitized video image
recorded 225 s after the start of the run. The aggregate is rough,
but, at least within the resolution of my imaging system, it does not
yet contain a large number of overhangs.  Curve (b) of Fig.
\ref{interfs} is the single-valued function $p_1(x)$ extracted from curve
(a); it is very similar to $p(x)$. Curve (c) shows $p(x)$ at a later
time, 385 s after the start of the run. By this time the aggregate is
more strongly branched and displays substantially more overhangs.
Curve (d) is the corresponding single-valued function $p_1(x)$.  In
this case there are significant differences between $p_1(x)$ and
$p(x)$ and Fourier analysis of $p_1(x)$ does not provide meaningful
information about the true interface.

By Fourier analysing the single-valued interface functions obtained
from a time-sequence of images, one can extract the Fourier power
$|A(k)|^2$ of a mode of wave number $k$ as a function of time. Fig.
\ref{powervst} shows $|A(k)|^2$ vs. time for three different values of
$k$, from the run illustrated in Fig. \ref{interfs}. In this
particular run the growth at the cathode first became visible about
100 s after the start of the run.  The Fourier power grows
exponentially with time between this time, and about 230 s. The dotted
lines in Fig.  \ref{powervst} are fits of the data to a growing
exponential over this time range. The growth rate depends on $k$.  At
times beyond the region of exponential growth, the growth rate
generally decreases. In this regime, however, $p_1(x)$ is not a good
approximation to the true interface and quantitative analysis is not
possible using this technique.  The interface shown in curves (a) and
(b) of Fig. \ref{interfs} corresponds to a time near the end of the
exponential growth phase of Fig. \ref{powervst}, while that of curves
(c) and (d) of Fig. \ref{interfs} is well beyond the end of this
phase.

Fig. \ref{dispersion} shows the exponential growth rate $\beta$ of the
Fourier {\it amplitude}\/ $|A(k)|$ as a function of $k$. The dotted
curve is a least-squares fit to Eq. \ref{MS}, which does not describe
the data well.  Rather, the growth rate increases more rapidly for
small $k$, displays a weak maximum, and then decreases more slowly for
larger $k$.  The solid curve plotted in Fig.  \ref{dispersion} is a
least-squares fit of the data to a dispersion relation with the form
of Eq. \ref{BMT}.  The fit is quite good, although, as in Fig.
\ref{dispersion}, most data sets show perhaps a bit more of a
peak in the data than the theoretical curve would indicate. Although
fits of the experimental data to Eq. \ref{BMT} looked satisfactory to
the eye, the uncertainties in the parameters were in many cases quite
large. This was due to the fairly large scatter in the $\beta(k)$
data; that in the data shown in Fig. \ref{dispersion} is typical.

{}From fits of Eq. \ref{BMT} to the data from a number of runs, we
determined the growth rate $\beta_{\hbox{\it max}}$ of the fastest
growing mode. Fig. \ref {beta}(a) is a plot of $\beta_{\hbox{\it
max}}$ against the cell current $I$ for two values of the
concentration $c$. Although the scatter in the data is considerably
larger than the error bars, the trend of the data is well described by
the power law $\beta_{\hbox{\it max}} \propto I^{\delta_1}$ with an
exponent $\delta_1 = 1.52 \pm 0.19$ for $c=0.02$ M and $\delta_1 = 1.27
\pm 0.06$ for $c = 0.1$ M. Fig. \ref{beta}(b) shows $\beta_{\hbox{\it
max}}$ as a function of $c$ for a fixed current. In this case a power
law fit to the data gives $\beta_{\hbox{\it max}}
\propto c^{-\delta_2}$, with $\delta_2 = 1.40 \pm 0.16$. The similarity
between the values of the exponents $\delta_1$ and $\delta_2$ suggests
that $\beta_{\hbox{\it max}}$ may be a function of the ratio $I/c$
alone.  In Fig. \ref{beta}(c) the same three data sets are plotted as
a function of $I/c$. With some scatter, the data collapse onto a
single power-law function given by $\beta_{\hbox{\it max}} = (5.2 \pm
1.2)\times 10^{-3} (I/c)^{1.37\pm0.08}$ s$^-1$.

To compare my results with the theoretical predictions \cite{BMT89}, I
make the assumption that my measurements were performed at currents
such that $i \ll i_L$, where $i$ is the current density and $i_L$ the
limiting current density \cite{electrochem}, and use bulk values for
the electrolyte properties involved in the expressions for the
coefficients $q$, $r$, and $s$
\cite{parameters}. The theory then predicts that the
coefficients will depend on current and concentration as $q \propto
I$, $r \propto c/I$, and $s \propto c/I$.  On the other hand, the
values of $q$ obtained from fits to the experimental data tend to
increase with $I$ at constant $c$, but decrease with $c$ at constant
$I$; the fitted values of $r$ show no systematic variation with either
$I$ or $c$ and in fact are constant to within a factor of two over the
range of conditions studied; and the values of $s$, while they have
large error bars, also show no systematic variation with either $I$ or
$c$.  Numerically, for $c = 0.1$ M and $I = 2 mA$, the value of $q$
obtained from fits to the experimental data is three orders of
magnitude larger than predicted, $r$ is a factor of 20 larger than
predicted, while $s$ is approxiamtely equal to the predicted value.
Under the same assumptions, the theory predicts that $\beta_{\hbox{\it
max}} \propto I^2/c$, while Fig. \ref{beta}(c) indicates that,
experimentally, $\beta_{\hbox{\it max}}$ behaves as a power law in
$I/c$.

The fact that the shape of the experimental dispersion relation can be
fitted by Eq. \ref{BMT} suggests that the basic ingredients of the
theory --- that is, that the instability is driven by mass transport,
stabilized by surface tension, and modified by electrochemical
processes --- are correct. However, the quantitative differences
between experiment and theory outlined above suggest that either the
assumptions made by me in calculating the coefficients are invalid, or
that effects not accounted for in the theory play an important role in
these experiments.

Both $q$ and $s$ are predicted to diverge when $i = i_L$
\cite{parameters}. In calculating the coefficients, I assumed that $i
\ll i_L$. A rough estimate of the limiting current for my cells
\cite{electrochem} gives $I_L
\approx 5$ mA for $c = 0.1$ M, similar to the highest currents used
at that concentration.  However, to get a thousand-fold increase in
$q$, one would need to be within 0.1\% of $i_L$. I find larger than
expected values of $q$ over a large range of currents, and the
assumption that $i \ll i_L$ is safe over most of that range. On the
other hand, it is quite likely that the use of the bulk value of the
electrolyte conductivity in evaluating the coefficients is a poor
approximation, since in an unsupported electrolyte a diffusion layer
depleted of ions, and with a correspondingly low conductivity,
develops near the cathode. However, using a conductivity of zero would
decrease $q$ by roughly a factor of two, while $r$ and $s$ would go to
zero \cite{parameters}. Thus these assumptions cannot account for the
observed discrepencies.

The theory of BMT assumes that ion transport is due to diffusion only.
In an unsupported binary electrolyte, migration will also be
important.  This can be accounted for by a correction factor which
would result in a decrease of the theoretical value of $q$ by about
20\% \cite{dpb95}.

The most likely explanation for the differences between my
experimental results and the theoretical predictions is the fact that
high electric fields in the diffusion layer cause electroconvective
flow in quasi-two-dimensional ECD experiments
\cite{fcr92,fcr93}. This flow has an influence on ion transport and on
the deposition process, and is not accounted for in the theory of BMT.
Natural convection, driven by density gradients in the electrolyte
\cite{rcfc94,bwlr94,d95,huth} may similarly play a role.

In summary, I have measured the growth rate of spatial modes as a
function of wave number during the early stages of the growth of metal
aggregates in thin-layer electrochemical deposition. The dispersion
relation is not well described by a MS-type theory \cite{ms},
but can be fitted to the form predicted by the stability analysis of
Ref. \cite{BMT89}. Significant quantitative differences between my
results and the predictions of the theory of Ref. \cite{BMT89} suggest
that convection in the electrolyte driven by electric fields or by
density gradients modifies the ion transport substantially and plays an
important role in determining the growth rate. A more detailed account
of this work will appear elsewhere \cite{tbp}.

This research was supported by the Natural Sciences and Engineering
Research Council of Canada. I am grateful to D. Barkey, G. Marshall,
G. White and S. Morris for helpful discussions.

\begin{figure}
\caption{The edge of the metal aggregate growing on the cathode
for a run with $c = 0.1$ M and $I = 1.4225$ mA. (a) The interface
function $p(x)$ at time $t = 225$ s, near the end of the exponential
growth regime seen in Fig.  \protect\ref{powervst} below. (b) The
single-valued function $p_1(x)$ corresponding to curve (a). (c) $p(x)$
at $t= 385$ s, well beyond the end of the exponential regime. (d) The
single-valued function $p_1(x)$ corresponding to curve (c). The
solid bars have length 1mm. }

\label{interfs}
\end{figure}

\begin{figure}

\caption{The Fourier power as a function of time for three values of wave
number $k$, for the same run as shown in Fig. \protect\ref{interfs}.
Triangles: $k = 27.2$ mm$^{-1}$, circles: $k = 75.3$ mm$^{-1}$, diamonds:
$k = 113$ mm$^{-1}$. The lines are fits to the data in the exponential
growth regime.}
\label{powervst}
\end{figure}

\begin{figure}

\caption{ The growth rate $\beta(k)$ of the Fourier amplitude $|A|$,
for the same run as the previous figures.  The dotted
curve is a fit to Eq. \protect\ref{MS}, and does not describe the data
well. The solid curve is a fit to the form predicted by BMT, Eq.
\protect\ref{BMT}.}

\label{dispersion}
\end{figure}

\begin{figure}
\caption{ (a) The maximum growth rate $\beta_{\hbox{\it max}}$ as a function of
the current $I$ for fixed concentration $c$. Squares: $c = 0.02$ M;
triangles: $c = 0.10$ M. The lines are fits of the data to
power laws in $I$, as discussed in the text. (b) $\beta_{\hbox{\it
max}}$ as a function of $c$ for $I = 2.00$ mA. The dashed line is a
fit to a power law. (c) The same data as in (a) and (b) plotted as a
function of $I/c$. The symbols are as in (a) and (b), and the
line is a fit of all the data to a power law.}
\label{beta}
\end{figure}


\begin{references}
\bibitem{parameters} From Eq. (41) of Ref.~ \protect\cite{BMT89},\\
$q = (\sqrt2 v\kappa/nF) \left[i_a/\kappa +
RTi/\left(n^2F^2Dc(1-i/i_L)\right)\right]$;\\
$qr = 2\sqrt2v^2\gamma\kappa/n^2F^2$;\\
and $s = (\sqrt2RT\kappa/nF)\left[1/\left(i_L(1 - i/i_L)\right) +
1/i\alpha\right]$; where $R$
is the gas constant, $T$ is the temperature, $F$ the Faraday constant,
$\gamma$ the surface energy per unit area of the deposited metal,
$\kappa$ the conductivity, $D$ the diffusion constant, $n$ the charge
on the cations, $\alpha$ the charge transfer coefficient, $i_L$ the
limiting current density of the cell, $i$ the actual current density,
$i_a$ the uniform current density with a flat electrode, and $c$ the
concentration.
\end{references}
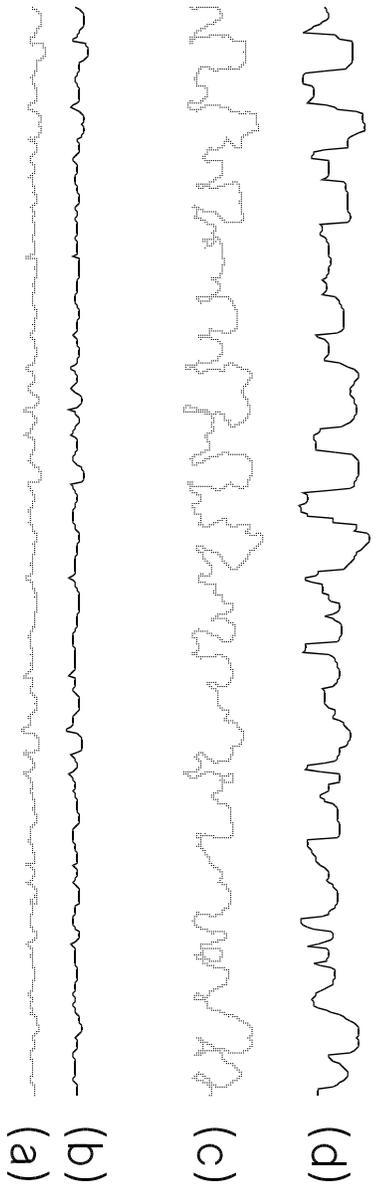
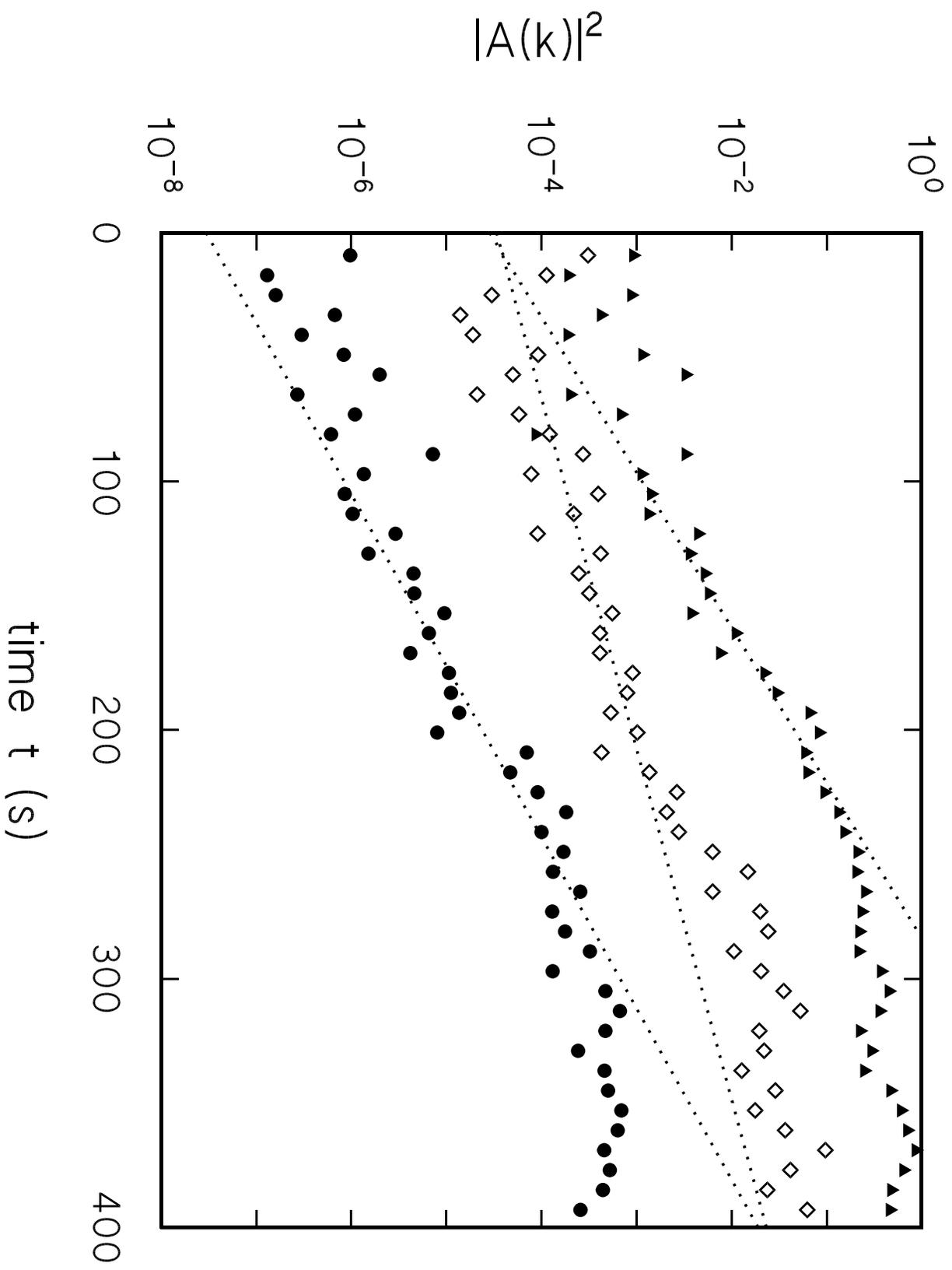
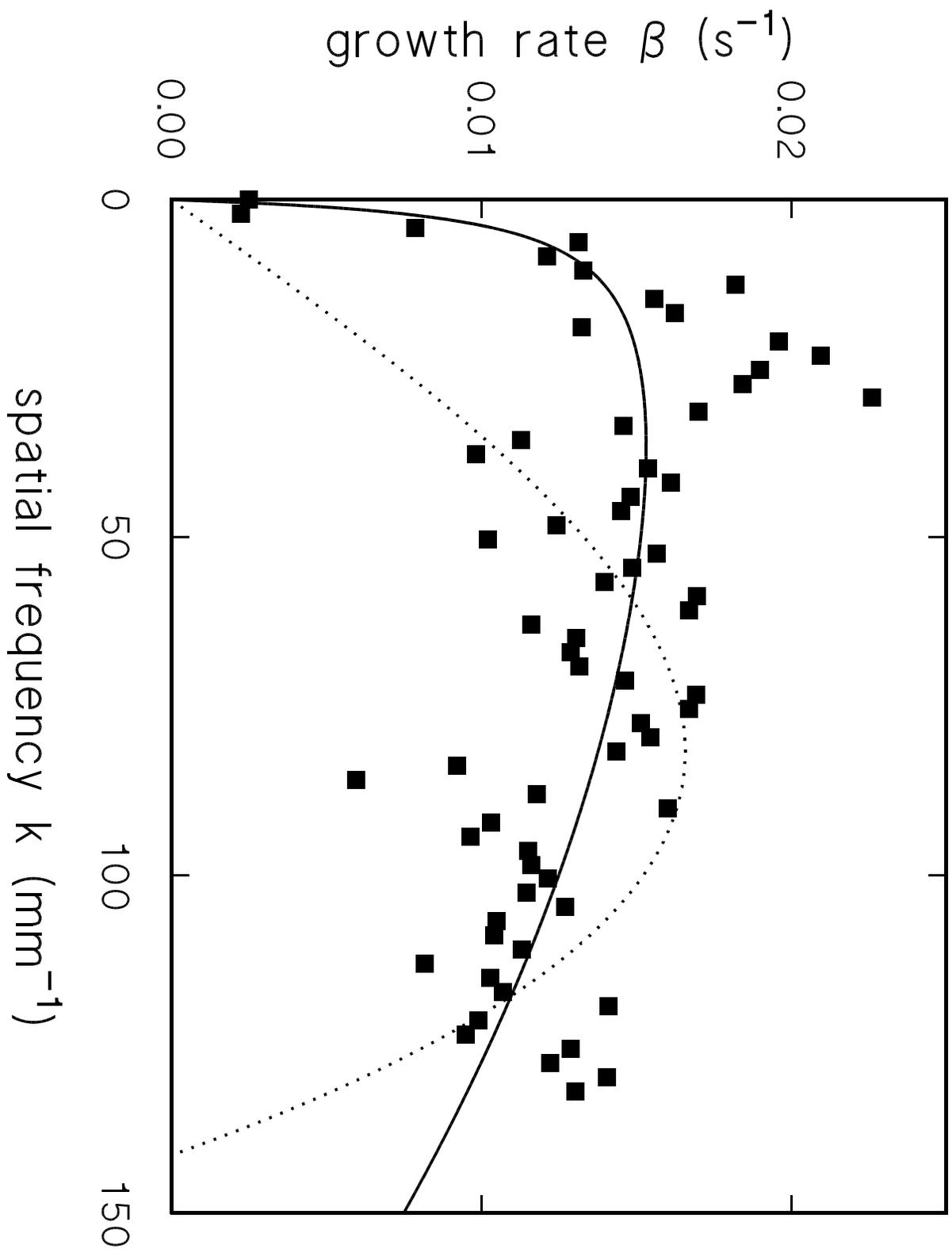
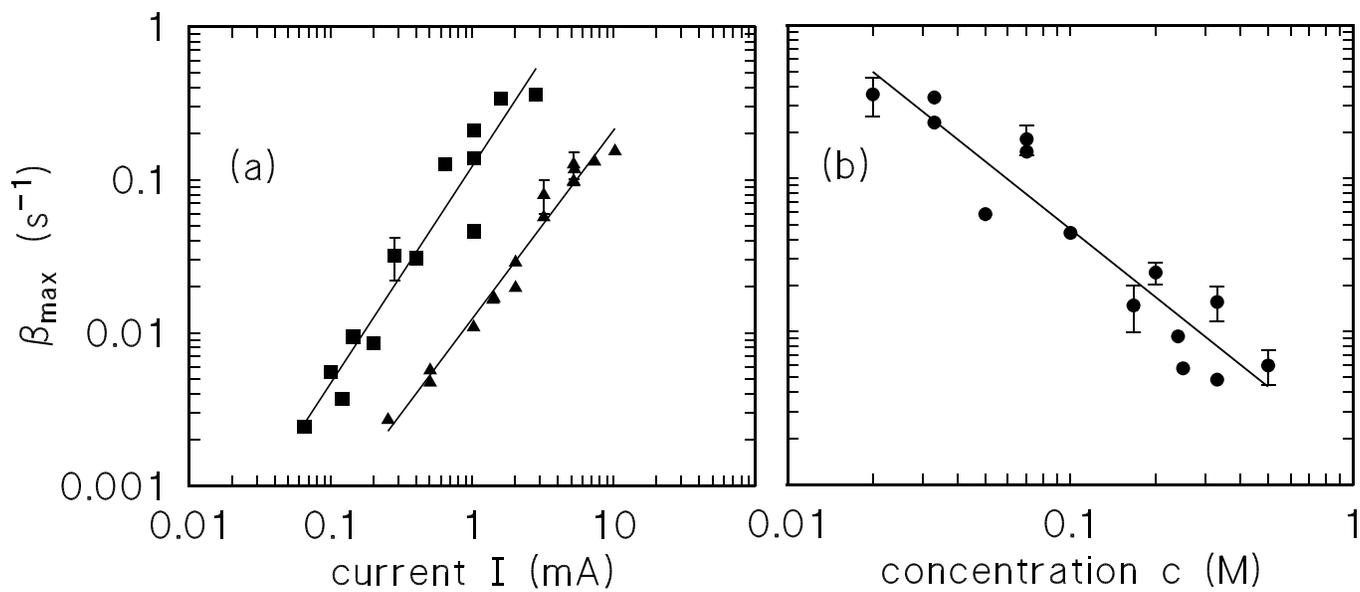
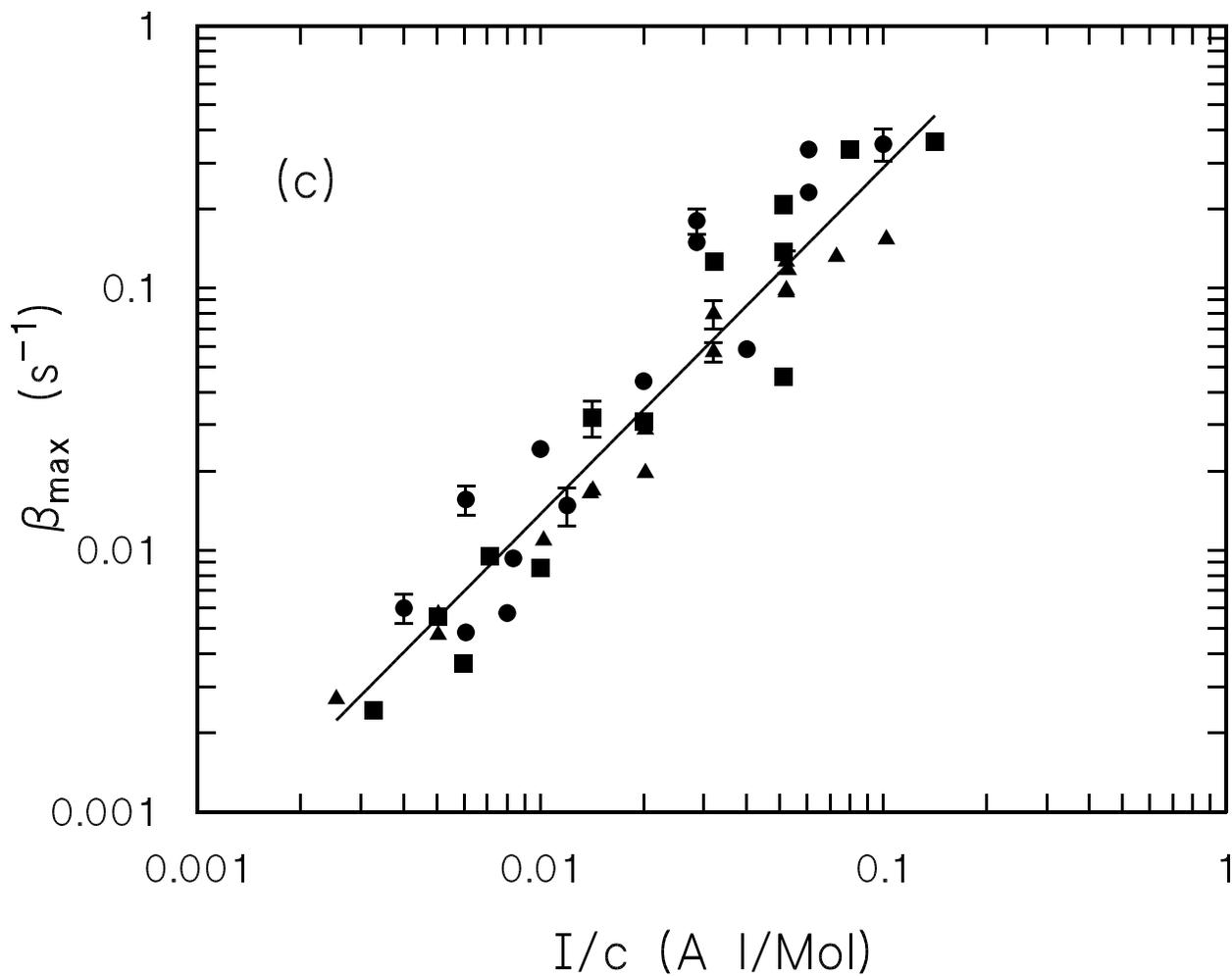
\end{document}